\documentclass[aps,prb,twocolumn,showpacs]{revtex4}
\usepackage[dvips]{graphicx}
\usepackage{amsmath}
\usepackage{times}

\begin{document}

\title{Nonlinear statistics of quantum transport in chaotic cavities}
\date{November 12, 2007} %
\author{D. V. Savin$^1$, H.-J. Sommers$^2$, and W. Wieczorek$^2$}
\affiliation{%
$^1$\,Department of Mathematical Sciences, Brunel University, Uxbridge UB8 3PH, UK \\
$^2$\,Fachbereich Physik, Universit\"at Duisburg-Essen, 47048 Duisburg, Germany %
}

\begin{abstract}
In the framework of the random matrix approach, we apply the theory of
Selberg's integral to problems of quantum transport in chaotic cavities. All
the moments of transmission eigenvalues are calculated analytically up to the
fourth order. As a result, we derive exact explicit expressions for the
skewness and kurtosis of the conductance and transmitted charge as well as
for the variance of the shot-noise power in chaotic cavities. The obtained
results are generally valid at arbitrary numbers of propagating channels in
the two attached leads. In the particular limit of large (and equal) channel
numbers, the shot-noise variance attends the universal value $1/64\beta$ that
determines a universal Gaussian statistics of shot-noise fluctuations in this
case.
\end{abstract}
\pacs{73.23.-b, 73.50.Td, 05.45.Mt, 73.63.Kv}

\maketitle

Quantum transport of non-interacting electrons in mesoscopic systems can be
conventionally described  in the framework of scattering
theory.\cite{Beenakker1997,Blanter2000} The conductance and shot-noise are
the brightest and, perhaps, most frequently considered examples.  Being
expressed in terms of transmission eigenvalues $T_i$ of a conductor, the
dimensionless conductance, $g$, and the zero-frequency shot-noise power, $p$,
of the two-terminal setup at zero temperature are respectively given by
\begin{equation}
  g = \sum_{i}^{n}T_i \quad\mathrm{and}\quad p = \sum_{i}^{n}T_i(1-T_i)\,.
\end{equation}
Here, $n\equiv\mathrm{min}(N_1,N_2)$ where $N_{1,2}$ is the number of
scattering channels in each of the two attached leads.

Generally, the positive quantities $T_i\leq1$ are mutually correlated random
numbers whose fluctuations depend on the conductor's nature. In the case of
chaotic cavities considered below, the joint distribution of $T_i$ is
believed to be provided by the random matrix theory (RMT) and reads as
follows\cite{Beenakker1997}
\begin{equation}\label{jpd}
  \mathcal{P}(\{T_i\}) = \mathcal{N}_\beta^{-1} \prod_{i=1}^n T_i^{\alpha -1}
  \prod_{j<k}|T_j - T_k |^\beta\,,
\end{equation}
with $\alpha=\frac{\beta}{2}(|N_1-N_2|+1)$ and normalization constant
$\mathcal{N}_\beta$. The symmetry index $\beta$ ($=1$, 2, or 4) distinguishes
between the three standard RMT classes (orthogonal, unitary, or symplectic
ensembles, respectively) which are realized depending on the presence or
absence of time-reversal and spin-flip symmetry in the
system.\cite{Beenakker1997,Mehta2}

The exact RMT results for the average and variance of the conductance are
well known for quite a long
time\cite{Beenakker1997,Baranger1994,Jalabert1994} and that for the average
shot-noise power has become available only recently.\cite{Savin2006} As
concerns higher order cumulants of these quantities, their exact RMT
expressions valid at arbitrary $N_{1,2}$ have not been reported in the
literature so far, a progress in this direction being announced very
recently.\cite{Sommers2007a}

In this communication, we answer this question by further developing and
applying the theory of Selberg's integral to the problem. As was recently
recognized,\cite{Savin2006} such an approach is a powerful non-perturbative
method suited particularly well for the studies of moments and counting
statistics, see also Refs.~7 and 8. It represents a useful alternative to
orthogonal polynomial\cite{Mehta2} or diagrammatic\cite{Brouwer1996}
approaches, especially in the situation when the finite dimensionality of
relevant random matrices becomes important.

Selberg's integral appears naturally in the problem first as an integral
determining the normalization constant
\begin{equation}\label{N}
  \mathcal{N}_\beta = \prod_{j =1}^{n-1} \frac{\Gamma(1+\frac{\beta}{2}(1+ j))
  \; \Gamma(\alpha+\frac{\beta}{2}j)\; \Gamma(1+\frac{\beta}{2}j) }{
  \Gamma(1+\frac{\beta}{2})\; \Gamma(1+\alpha+\frac{\beta}{2}(n+j-1))}\,.
\end{equation}
This expression assures that (\ref{jpd}) is a probability density, being
generally valid for discrete  $n$ and continuous $\alpha$ and
$\beta$.\cite{Selberg} To study the cumulants of $g$ and $p$, one needs to
know the moments $\langle T_1^{n_1} \cdots T_k^{n_k}\rangle$, with
$\langle\ldots\rangle$ standing for the integration over the joint
probability density (\ref{jpd}) and $n_i\geq0$. Here, we calculate all the
moments with $\sum_i n_i \leq 4$ by deriving a set of algebraic relations for
them and reducing the moments to forms of Selberg's integral. Presenting the
relevant technical details at the end of the paper, we now discuss
applications of the obtained results to various linear and nonlinear
statistics on the transmission eigenvalues.

\emph{Transmitted charge cumulants.}-- The statistics of charge $q$ (in units
of $e$) transmitted through the cavity over the observation time is usually
described by means of the current or charge cumulants,
$\langle\langle{q^m}\rangle\rangle$. Following Levitov et
al.\cite{Levitov1993,Lee1995}, it is convenient to use a general formula for
the cumulant generating function expressed in terms of the transmission
eigenvalues as follows
$\sum_m\frac{\lambda^m}{m!}\langle\langle{q^m}\rangle\rangle =
\langle\sum_i\ln[1+T_i(e^{\lambda}-1)]\rangle$. One finds that the first
cumulant, $\langle\langle{q}\rangle\rangle=n\langle{T_1}\rangle$, gives the
conductance, the RMT average of which is known \cite{Baranger1994} to be
\begin{equation}\label{<g>}
  \langle\langle q \rangle\rangle \equiv \langle g \rangle = \frac{N_1 N_2}{N
  -1 + \frac{2}{\beta}}\,, \qquad N\equiv N_1+N_2\,,
\end{equation}
while
$\langle\langle{q^2}\rangle\rangle=n(\langle{T_1}\rangle-\langle{T_1^2}\rangle)$
yields shot-noise as follows\cite{Savin2006}
\begin{equation}\label{<p>}
  \langle\langle q^2 \rangle\rangle \equiv \langle p \rangle = \langle g
  \rangle \frac{(N_1-1+\frac{2}{\beta})(N_2-1+\frac{2}{\beta}) }{
(N-2+\frac{2}{\beta})(N-1+\frac{4}{\beta})}\,.
\end{equation}
An equivalent to RMT derivation of these and related results within a
semiclassical approach may be found in Ref.~13.

The charge cumulants are an example of linear statistics on the $T_i$'s that
is fully determined by the transmission eigenvalue density,
$\rho(T)=\langle\sum_i\delta(T-T_i)\rangle$. However, the latter is
analytically known only in some limiting cases of a few
\cite{Baranger1994,Araujo1998} or many \cite{Nazarov1995} open channels,
restricting the use of $\rho(T)$ for the calculation of
$\langle\langle{q^m}\rangle\rangle$ and full counting
statistics\cite{Blanter2001i,Bulashenko2005} to these cases.

In the general situation of arbitrary $N_{1,2}$, one can alternatively
consider the joint probability distribution (\ref{jpd}) and exploit its
simple algebraic structure (i.e. that of the Selberg's integral kernel) to
derive exact relations for its moments.\cite{Savin2006} For example, Eqs.
(\ref{Pi_m}) and (\ref{T^2}) presented below yield straightforwardly and in a
uniform way exact results (\ref{<g>}), (\ref{<p>}) and (\ref{varg}) for
$\langle{g}\rangle$, $\langle{p}\rangle$ and $\mathrm{var}(g)$, respectively.
This approach was recently extended further to find the third cumulant
$\langle\langle{q^3}\rangle\rangle=\langle(q-\langle{q}\rangle)^3\rangle$
exactly.\cite{Novaes2007} For completeness and later use, we write down this
result (following from Eq.~(\ref{T^3}) below) as follows
\begin{eqnarray}\label{q^3}
 \frac{\langle\langle q^3 \rangle\rangle}{\mathrm{var}(q)} =
 \frac{(1-\frac{2}{\beta})^2-(N_1-N_2)^2}{(N-3+\frac{2}{\beta})(N-1+\frac{6}{\beta})}\,.
\end{eqnarray}

As to the fourth cumulant of the transmitted charge, its explicit expression
can be found from Eqs.~(\ref{Pi_m}) -- (\ref{T^4}) according to
$\langle\langle{q^4}\rangle\rangle = n[\langle{T_1}\rangle -
7\langle{T_1^2}\rangle + 12\langle{T_1^3}\rangle - 6\langle{T_1^4}\rangle$],
the result being too lengthy to be reported here. In the case of the
single-mode leads, $N_1=N_2=1$, one gets $\langle\langle
q^4\rangle\rangle=-\frac{2}{105}$, $-\frac{1}{30}$, $-\frac{1}{30}$ at the
values of $\beta=1$, 2, 4, respectively. In the opposite semiclassical limit
of many channels, $N_{1,2}\gg1$, we represent the outcome of our calculation
as the following $\frac{1}{N}$ expansion:
\begin{eqnarray}
\label{q^4}
 \frac{\langle\langle q^4\rangle\rangle}{\mathrm{var}(q)} &=&
 \frac{N_1^4 - 8N_1N_2^3 + 12N_1^2N_2^2 - 8N_1^3N_2 + N_2^4}{N^4}
 \nonumber \\
 && + \frac{6(\beta-2)(N_1-N_2)^2(2N_1^2 - 7N_1N_2 + 2N_2^2)}{\beta N^5}
 \nonumber \\
 && + {\cal O}(1/N^2)\,.
\end{eqnarray}
The leading order term agrees with the result obtained by a different method
in Ref.~\cite{Blanter2001i} The next order term gives a week localization
correction which vanishes at $\beta=2$ or $N_1=N_2$. In the latter case of
symmetric cavities, one further finds
\begin{equation}
 \langle\langle q^4\rangle\rangle = \frac{n}{64}
 \biggl( 1-\frac{\beta^2-6\beta+4}{2\beta^2n^2} \biggr)
 + \mathcal{O}\Bigl(\frac{1}{n^2}\Bigr)\,.
\end{equation}

\emph{Conductance cumulants.}-- As is clear from the discussion, the
presented method is equally applied to nonlinear statistics determined by
different transmission eigenvalues as well. The simplest example of such a
quantity is the variance of the conductance, the exact RMT result of which
reads\cite{Beenakker1997,Savin2006}
\begin{equation}\label{varg}
  \frac{\mbox{var}(g)}{\langle{g}\rangle} =
  \frac{2(N_1-1+\frac{2}{\beta})(N_2-1+\frac{2}{\beta}) }{
  \beta(N-2+\frac{2}{\beta})(N-1+\frac{2}{\beta})(N-1+\frac{4}{\beta})} \,.
\end{equation}
We note that $\mathrm{var}(g)=2\langle{g}\rangle\langle{p}\rangle/\beta
N_1N_2$ makes a relation of (\ref{varg}) to the linear statics (\ref{<g>})
and (\ref{<p>}) considered above.

The third cumulant of the conductance, the so-called skewness, can be also
found from Eqs.~(\ref{Pi_m}) -- (\ref{T^3}) and be represented after some
algebra in the following compact form:
\begin{eqnarray}
 \frac{\langle \langle g^3 \rangle \rangle}{\mathrm{var}(g)}
 = \frac{4[(1-\frac{2}{\beta})^2-(N_1-N_2)^2] }{
   \beta(N-3+ \frac{2}{\beta}) (N-1+\frac{2}{\beta})(N-1+\frac{6}{\beta})}\,.
\end{eqnarray}
One finds immediately that
$\langle\langle{g^3}\rangle\rangle=8[\langle{g}\rangle/\beta
N_1N_2]^2\langle\langle{q^3}\rangle\rangle$.  It is also worth noting that
the skewness vanishes for symmetric cavities ($N_1=N_2$) at $\beta=2$. This
holds generally for any odd cumulant of $g$ (or $q$), as the corresponding
distribution becomes symmetric around $\frac{n}{2}$ (or $\frac{1}{2}$) in
this case.\cite{symmetry}

By our method, we have also calculated the fourth cumulant
$\langle\langle{g^4}\rangle\rangle$ which is related to the conductance
kurtosis. The fourth moment of the conductance is found to be determined by
moments of $T_i$'s as $\langle{g^4}\rangle=n[\langle{T_1^4}\rangle +
(n-1)(3\langle{T_1^2T_2^2}\rangle + 4\langle{T_1T_2^3}\rangle) +
6(n-1)(n-2)\langle{T_1T_2T_3^2}\rangle +
(n-1)(n-2)(n-3)\langle{T_1T_2T_3T_4}\rangle]$, the corresponding cumulant
being given by the standard formula. Since the resulting explicit expression
appears to be too cumbersome, we restrict our consideration to the limiting
cases of the single-mode and many-mode leads. In the former case, we get
$\langle\langle g^4 \rangle\rangle=-\frac{32}{4725}$, $-\frac{1}{120}$,
$-\frac{1}{540}$ at $\beta=1$, 2, 4, respectively, whereas in the latter case
of $N_{1,2}\gg1$ we arrive at the following expression:
\begin{eqnarray}
\label{<g^4>}
 \frac{\langle\langle g^4\rangle\rangle}{\mathrm{var}(g)} &=&
 \frac{24 }{\beta^2 N^6}\Bigl[(N_1-N_2)^2 (N_1^2 + N_2^2 - 4 N_1 N_2)
 \nonumber \\
 && + \frac{\beta-2}{\beta N} \Bigl( 12(N_1^4 + N_2^4) -
 64N_1N_2(N_1^2 + N_2^2)
 \nonumber \\
 && + 105 N_1^2 N_2^2 \Bigr) \Bigr] + {\cal O} (1/N^4)\,.
\end{eqnarray}

One can readily see that higher cumulants contribute in the next order of
$\frac{1}{N}$, thus the conductance distribution gets more Gaussian-like as
$N$ grows.\cite{Altshuler1986b} This tendency becomes even stronger for
symmetric cavities at $\beta=2$, as then
$\langle\langle{g^3}\rangle\rangle=0$ identically and
$\langle\langle{g^4}\rangle\rangle$ vanishes in the leading and
next-to-leading orders. In this case of $N_{1,2}=n\gg1$, one gets
\begin{equation}
 \langle\langle g^4\rangle\rangle = \frac{3}{128\beta^3n^3}
 \biggl(1-\frac{2}{\beta} + \frac{(\beta+2)^2}{2\beta^2n}\biggr)
 + \mathcal{O}\Bigl(\frac{1}{n^5}\Bigr). \mbox{\qquad}
\end{equation}

\emph{Shot-noise variance.}-- Now we discuss statistics of the shot-noise
power. Its average value is given by (\ref{<p>}), whereas its second
cumulant, the variance, is determined by $\mathrm{var}(p) = n
[\langle{T_1^2}\rangle - 2\langle{T_1^3}\rangle + \langle{T_1^4}\rangle] +
n(n-1) [\langle{T_1T_2}\rangle - 2\langle{T_1T_2^2}\rangle +
\langle{T_1^2T_2^2}\rangle] -
n^2[\langle{T_1}\rangle-\langle{T_1^2}\rangle]^2$. However, as in all other
cases of the fourth order cumulants considered above, the explicit result for
$\mathrm{var}(p)$ can not be cast in a compact form, so we present the
limiting cases again. In the single-mode case, we get
$\mathrm{var}(p)=\frac{4}{525}$, $\frac{1}{180}$, $\frac{1}{180}$
correspondingly at $\beta=1$, 2, 4. In the many channel case, we obtain the
following expansion:
\begin{eqnarray}\label{varp}
 \frac{\mathrm{var}(p)}{\langle p \rangle} &=&
 \frac{2}{\beta N^5} \Bigl[ N_1^4 + N_2^4 - 4 N_1 N_2 (N_1-N_2)^2
 \nonumber \\
 && + \frac{\beta-2}{\beta N} \Bigl( 9(N_1^4 + N_2^4) -
 42N_1N_2(N_1^2 +  N_2^2)
 \nonumber \\
 && + 70N_1^2 N_1^2 \Bigr) \Bigr] + {\cal O} (1/N^3)\,.
\end{eqnarray}
As usual, the first weak localization correction vanishes for unitary
symmetry, $\beta=2$. The next order correction can be also found and reads as
follows
\begin{equation}
 \mathrm{var}(p) \approx \frac{1}{64\beta} \biggl(1 + \frac{\beta-2}{\beta n}
 + \frac{4+\beta(\beta-2)}{2\beta^2n^2}\biggr) \,,
\end{equation}
where we have omitted the terms $\sim{n}^{-3}$ and put $N_1=N_2$ for
simplicity. The general dependence of the shot-noise variance on the channel
numbers in the leads is illustrated on Fig.~1.
\begin{figure}[t]
  \includegraphics[width=0.45\textwidth]{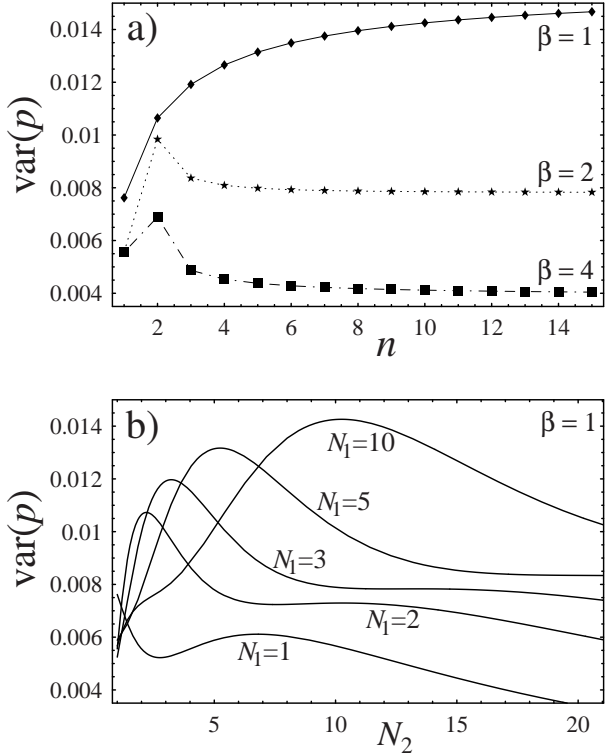}
  \caption{The variance of the shot-noise power in chaotic cavities as a function
   of the channel numbers in the leads:
a) The case of symmetric cavities ($N_1=N_2=n$) for three RMT ensembles, when
   $\mathrm{var}(p)$ saturates at the universal value $(64\beta)^{-1}$ at large $n$.
b) The case of asymmetric cavities with preserved time-reversal symmetry
   ($\beta=1$) for fixed number $N_1$ of channels in the one lead and varied one
   $N_2$ in the other lead. The shot-noise variance shows a well-developed maximum
   at $N_2\approx N_1$ and then diminishes down to zero according to
   $\mathrm{var}(p)\approx2N_1(N_1-1+\frac{2}{\beta})/\beta N_2^2$ as $N_2$ grows.}
\end{figure}

\emph{Covariance of $g$ and $p$.}-- It is also instructive to consider
statistical correlations between the conductance and the shot-noise power
that can be characterized by their covariance
$\mathrm{cov}(g,p)=\langle{gp}\rangle-\langle{g}\rangle\langle{p}\rangle$.
This quantity involves moments of $T_i$'s up to the third order.
Surprisingly, the resulting exact formula for
$\mathrm{cov}(g,p)/\mathrm{var}(g)$ turns out to be given precisely by the
r.h.s. of Eq.~(\ref{q^3}). This expression vanishes in symmetric cavities at
$\beta=2$ that can be easily seen again as a consequence of the symmetry of
the distributions. In this case, therefore, $g$ and $p$ become uncorrelated
on the level of their averages and it would be interesting to understand to
which extent this holds for their higher moments in general.

\emph{Moments of \{$T_i$\} and Selberg's integral.}-- Finally, we discuss the
derivation of general moments $\langle T_1^{n_1} \cdots T_k^{n_k}\rangle$ at
arbitrary positive $\alpha$ and $\beta$. Moments with all $n_i = 1$ as well
as $\langle T_1^2\rangle$ can be found from recursion relations already given
in Mehta's book \cite{Mehta2} which read as follows
\begin{equation}\label{Pi_m}
 \Pi_m \equiv \langle{T_1T_2\cdots T_m}\rangle = \prod_{j=1}^m
 \frac{\alpha+\frac{\beta}{2}(n-j)}{\alpha+1+\frac{\beta}{2}(2n-j-1)}\,,
\end{equation}
\begin{equation}\label{T^2}
 \langle{T_1^2}\rangle = \frac{[\alpha+1+\beta(n-1)]\Pi_1 -
 \frac{\beta}{2}(n-1)\Pi_2}{\alpha+2+\beta(n-1)}\,.
\end{equation}
To calculate moments containing higher powers of $T_m$, one may note that
$$
 \langle T_1 T_2 \ldots T_m^k \rangle = \langle T_1 T_2 \ldots
 \frac{\partial}{\partial T_m} \frac{T_m^{k+1}}{k+1} \rangle
$$
and employ a partial integration here (for $\alpha>0$). This yields a
contribution $\langle T_1T_2 \ldots \rangle'_{n-1}$ at the upper boundary
$T_m=1$, where notation $\langle\ldots\rangle'_{n-1}$ has been introduced for
an averaging over a joint density of eigenvalues $T_1,\ldots, T_{n-1}$ which
contains in addition a factor $\prod_{i=1}^{n-1}|1-T_i|^\beta$. This case is
also contained in the general form of Selberg's integral. In particular, the
corresponding analogue of (\ref{Pi_m}) is found to be
\begin{equation} \label{Pi'_m}
\hspace*{-1ex}\Pi'_m \equiv \langle{T_1\cdots T_m}\rangle'_{n-1} =
 \prod_{j=1}^m \frac{\alpha+\frac{\beta}{2}(n-j-1)}{\alpha+1+\frac{\beta}{2}(2n-j-1)}.
\end{equation}
In this way we are able to calculate all the moments up to the fourth order.
For moments of the third order we have
\begin{subequations}
\label{T^3}
\begin{eqnarray}
  && \langle T_1^3\rangle = \frac{\alpha+\frac{\beta}{2}(n-1)
  \left(1 - 2\langle T_1 T_2^2\rangle\right) }{ \alpha+3+\beta(n-1)}\,,
\\
  && \langle T_1 T_2^2\rangle = \frac{[\alpha+\beta(n-1)]\Pi'_1
  - \frac{\beta}{2}(n-2)\Pi_3 }{ \alpha+2+\beta(n-1)}\,.\mbox{\qquad}
\end{eqnarray}
\end{subequations}
The fourth order moments are given as follows
\begin{subequations}
\label{T^4}
\begin{eqnarray}
 && \langle T_1^4 \rangle =  \frac{\alpha + \frac{\beta}{2}(n-1)
  \left(1 - 2\langle T_1 T_2^3\rangle - \langle T_1^2 T_2^2\rangle\right)
  }{ \alpha+4+\beta(n-1) }\,,\mbox{\qquad}
\\
 && \langle T_1 T_2^3 \rangle = \frac{[\alpha + \frac{\beta}{2}(n-1)]\Pi'_1
  - \frac{\beta}{2} \langle T_1^2 T_2^2\rangle }{ \alpha+3+\beta(n-1) }
  \nonumber \\
 && \hspace*{10ex}
  - \frac{\beta(n-2)\langle T_1 T_2 T_3^2 \rangle }{ \alpha+3+\beta(n-1) }\,,
\\
 && \langle T_1^2 T_2^2\rangle = \frac{[\alpha + \frac{\beta}{2}(n-1)]
  \langle T_1^2\rangle'_{n-1} }{ \alpha+2+\beta(n-\frac{3}{2}) }
  \nonumber \\
  &&\hspace*{10ex} - \frac{\beta}{2}\frac{(n-2)\langle T_1 T_2 T_3^2\rangle
  }{ \alpha+2+\beta(n-\frac{3}{2}) }\,,
\\
 && \langle T_1 T_2 T_3^2\rangle = \frac{[\alpha+\frac{\beta}{2}(n-1)]\Pi'_2
  - \frac{\beta}{2}(n-3) \Pi_4 }{ \alpha+2+\beta(n-1)}\mbox{\qquad}
\end{eqnarray}
\end{subequations}
and
\begin{equation}
 \langle T_1^2 \rangle'_{n-1} = \frac{[\alpha+1+\beta(n-2)]\Pi'_1
  - \frac{\beta}{2}(n-2)\Pi'_2 }{ \alpha+2+\beta(n-1) }\,.
\end{equation}
At last, expressions (\ref{Pi_m}) and (\ref{Pi'_m}) taken at $m=1,\ldots,4$
make the above algebraic system of equations be closed.

\emph{Discussion.}-- We have applied essentially the theory of Selberg's
integral to problems of quantum transport in chaotic cavities. The cumulants
up to the fourth order for current and conductance fluctuations and up to the
second order for shot-noise have been calculated exactly at arbitrary channel
numbers and symmetry parameter $\beta$. We note that the proposed method can
be also used for determining the corresponding distributions in closed form
suitable for an analytic work in the case of a few open channels as well as
for numerical implementations.\cite{Sommers2007a,Wieczorec} Our analysis of
shot-noise statistics suggests that in close analogy with universal
conductance fluctuations, \cite{Altshuler1986b} we may characterize universal
fluctuations of shot-noise by their cumulants as follows
\begin{equation}\label{usnf}
\langle\langle{p^m}\rangle\rangle \propto \langle{p}\rangle^{2-m}\,.
\end{equation}
In the limit of large (and equal) channel numbers, Eq.~(\ref{usnf}) yields
the Gaussian distribution which is peaked at
$\langle{p}\rangle\approx\frac{n}{4}$ and has a width given by the universal
value of $\mathrm{var}(p)\approx\frac{1}{64\beta}$.  It would be highly
interesting to understand how our findings, which are relevant to the
zero-dimensional (RMT) case, can be extended to higher dimensions where some
analytical results are already known.\cite{Macedo1994b,vanRossum1997}

It would be also desirable to compare our results with the relevant
experimental data. For example, measurements of up to the fifth cumulants of
the transmitted charge have been recently done in a weakly coupled quantum
dot.\cite{Gustavsson2007} However, performing similar experiments for a lower
impedance device, such as a perfectly open chaotic cavity with several modes
in the leads, requires a much higher detector resolution that represents a
current experimental challenge. On the other hand, the conductance
distribution in chaotic cavities\cite{Brouwer1997ii} has been also tested by
means of experiments with microwaves.\cite{Hemmady2006b} A direct comparison
with our results in this case could be, however, not so straightforward, as
it involves taking into account effects of dephasing and absorption
\cite{Brouwer1997ii,Fyodorov2005r} as well.

As another potential application of our results, we mention quantized
transport in graphene \emph{p-n} junctions that has been very recently
studied both experimentally\cite{Williams2007} and
theoretically.\cite{Abanin2007} The exact RMT results for higher cumulants of
the conductance and noise may be useful for understanding possible mechanisms
of edge mode mixing in the bipolar regime leading to chaotic transport
there.\cite{Abanin2007} Further work in this and in the other directions
mentioned above is needed.

We thank K. Ensslin for correspondence and drawing our attention to Ref.~23.
The financial support by SFB/TR 12 der DFG and BRIEF grant is acknowledged
with thanks.


\end{document}